\documentclass[fleqn,usenatbib]{mnras}

\usepackage[T1]{fontenc}
\usepackage{ae,aecompl}

\usepackage{graphicx}
\usepackage{amsmath}
\usepackage{amssymb}
\usepackage{color}

\newcommand\rg{GM/c^2}
\newcommand\tg{GM/c^3}
\newcommand\mdot{\dot{M}}
\newcommand\mdotEdd{\dot{M}_{\rm Edd}}
\newcommand\mdotedd{\dot{m}}
\newcommand\msun{\rm M_{\rm \odot}}

\newcommand\mdotu{\rm [M_\odot yr^{-1}]}
\newcommand\mbh{M_{\rm BH}}
\newcommand\rhigh{R_{\rm high}}

\title[Scale-invariance of black                                                                                                                                   
  hole accretion]{Scale-invariance of black
  hole accretion: modeling emission from a black hole X-ray binary with
  relativistic accretion flow simulations}

\author[M. Mo{\'s}cibrodzka]{
M. Mo{\'s}cibrodzka$^{1}$\thanks{E-mail: m.moscibrodzka@astro.ru.nl}\\
Department of Astrophysics/IMAPP, Radboud University, P.O. Box 9010, 6500 GL
Nijmegen, The Netherlands
}

\date{Accepted XXX. Received YYY; in original form ZZZ}

\pubyear{2019}

\begin{document}
\label{firstpage}
\pagerange{\pageref{firstpage}--\pageref{lastpage}}
\maketitle

\begin{abstract}
We model non-thermal emission spectrum of the extremely sub-Eddington X-ray
binary system A0620-00. It is believed that this non-thermal emission is
produced by a radiatively inefficient "quiescent" accretion onto a stellar-mass
black hole present in the system.  We post-process GRMHD simulations with
multiwavelength, fully polarized, relativistic radiative transfer calculations
to predict broadband spectra and emission polarization levels for a range of electron
models and accretion rates. We find that a model with strong coupling of
electrons and ions in the accretion disk and accretion rate of only
$\dot{M}=3\times10^{-13} \mdotu$ is able to recover the observed
X-ray spectral slope as well as the excess of linear polarization detected in
the source in NIR/optical bands. Our models constrain the spectral
properties of a putative relativistic jet produced in this system.
In addition we show that the magnetized winds from our
hot accretion flow carry away a small fraction of the orbital angular momentum
of the binary unable to explain the observed rapid orbital decay of the system. Similar
to the present GRMHD simulations are often used to explain emission from sub-Eddington
supermassvie black holes in Sgr~A* or M87; the present simulations
allow us to test whether some aspects of the quiescent black hole accretion
are scale invariant. 
\end{abstract}


\section{Introduction}\label{sect:intro}

A0620-00 is a low mass X-ray binary system composed of a main sequence
K-type star (V616 Mon) in orbit around a compact object. The orbital period of
the system $P \approx 7.7$ hr derived from variations in optical light curves
yields the separation between the two components $\sim 3-5 R_{\odot}$. The
ellipsoidal shape of the optical light curve from the stellar companion
indicates the tidal deformation of the star \citep{mcclintock:1986}; the shape
of optical light-curves are used to model the mass of the companion and the
inclination of the system with respect to our line of
sight. \citealt{cantrell:2008} carried out
a comprehensive analyses of 30 years of optical observations of this system
and estimated the mass of the compact component to be $\mbh = 6.6\pm0.25 \msun$
strongly favoring a black hole and the inclination of the system is
$51\pm0.9 deg$ (however, see e.g., \citealt{shahbaz:1994},
\citealt{vGrunsven:2017} for alternative evaluations).  Notice that the
precise determination of $\mbh$ and $i$ depends on the uncertain details of
the optical emission produced by a putative accretion disk around the accretor
and emission from a hot spot at the outer edge of the disk
\citep{neilsen:2008}. Above $\mbh$ and $i$ yield distance to the system to
be $D=1.06 \pm 0.12$kpc, making A0620-00 the closest currently known (stellar-mass) black
hole system to Earth \citep{cantrell:2010}.

These detailed studies of A0620-00 followed X-ray nova outburst in the source
in 1975 when the source become the brightest X-ray source on the sky for a few
days ($L_{\rm X} \approx 10^{38}{\rm [ergs/s]}$, \citealt{elvis:1975}).  Since
the outburst the source X-ray luminosity of A0620-00 is extremely low $L_{\rm
  X} \leq 10^{31}$ \citep{kong:2002} and the system is detectable in X-rays
due to its proximity. In units of Eddington luminosities (where Eddington
luminosity is defined as usually: $L_{\rm Edd} \equiv 4 \pi G M_{BH} m_p
c/\sigma_{\rm TH} = 1.26 \times 10^{38} (\mbh/M_{\odot}) \,\, {\rm [ergs \,\,
    s^{-1}]}$),
A0620-00 is a strongly sub-Eddington source, $L_{\rm X} \leq 10^{-7} L_{\rm Edd}$ \citep{dincer:2018} that is usually classified as
a quiescent phase of accretion. In this phase the accretion disk around the
black hole is believed to take form of a radiatively inefficient/advection
dominated flow (RIAF/ADAF, \citealt{yuan:2014}).  There are only a few other known black hole
systems in strongly sub-Eddington state: XTE
J1118+480 ($L_{\rm X}/L_{\rm Edd}=4 \times 10^{-9}$, \citealt{gallo:2014}),
Swift J1357.2-0933 ($L_{\rm X}/L_{\rm Edd}=4 \times 10^{-9} - 3 \times 10^{-8}$, \citealt{armas:2014,plotkin:2016}), and Sgr~A* associated
with supermassvie black hole in the center of our Galaxy ($L_{\rm X}/L_{\rm Edd} < 10^{-10}$, \citealt{baganoff:2003}).
Besides optical/NIR and X-ray monitoring, A0620-00 system is frequently observed in radio \citep{gallo:2006}
and recently Atacama Large Millimeter Array has detected the millimeter
counterpart of the system \citep{gallo:2019}.

The most prominent component of A0620-00 multiwavelenght spectrum is a
optical/NIR thermal hump produced predominantly by the companion star with
some contribution of non-thermal emission from accretion processes. The
non-stellar emission in radio/mm is believed to originate in a relativistic
outflow/jet. The X-ray emission is thought to be produced by synchrotron
process (via direct synchrotron or self-synchrotron Compton emission) at the
jet base and/or by a RIAF/ADAF launching the jet. The details of the jet
structure and the jet launching region, the strength of magnetic fields and
particle acceleration, are constrained only in terms of simplified
semi-analytic models (e.g., \citealt{plotkin:2015,connors:2017}).

In this work, we use a global model of relativistic hot magnetized accretion
disk to model non-thermal emission from A0620-00 system. The multiwavelenght
emission spectra are calculated using selfconsistent models of accretion
realized via three dimensional general relativistic
magnetohydrodynamics (3D GRMHD) numerical simulations and self-consisitent
general relativistic radiative transfer models. The main motivation for
discussing the accretion onto black holes in X-ray binary system in context of
a more detailed (compared to all previous ones) models are the recent results
from the Event Horizon Telescope (EHT). The telescope mapped
a hot accretion flow around sub-Eddington supermassive
black hole in M87 galaxy on horizon scales \citep{ehtIV:2019} and
GRMHD models of radiatively inefficient accretion
have been used to interpret the observations \citep{ehtV:2019}. The major
uncertainty in these GRMHD models was the
thermodynamics of electrons that allowed various different physical scenarios
to be acceptable. The same uncertainty concerns future modeling of
Sgr~A*, which is the second target for EHT.
It is believed that black holes of stellar and
supermassive masses accrete and produce jets in the same way at a given
accretion rate scaled to their Eddington limit (\citealt{merloni:2003},
\citealt{falcke:2004}). Accretion flow around the stellar-mass black hole in
A0620-00 is not only another ideal source for observational tests of
electron thermodynamics in GRMHD models that may help us to understand the general
structure of strongly sub-Eddington flows but also, together with EHT results, allows us to test
and understand the similarities of physical processes nearby supermassvie and stellar-mass black
hole horizons.

The structure of the paper is as follows.  In Sect.~\ref{sect:model} we
describe details of our numerical modeling techniques and adopted physical
scenario of accretion. In Sects.~\ref{sect:results}
and \ref{sect:moreresults}, we present the results of
the modeling. We discuss the results in
Sect.~\ref{sect:discussion}.

\section{Model of polarized multiwavelenght emission from accreting black hole}\label{sect:model}

In quiescent state, the dynamics of RIAF/ADAF and putative jet is
unaffected by radiative losses \citep{ryan:2017}.
Hence A0620-00 can be considered in terms of a non-radiative GRMHD
flow models.

We integrate the equations of GRMHD starting with the following
  initial and boundary conditions. Our initial distribution of plasma rest-mass
density and internal energy is described by an analytical model of a torus in
Keplerian orbit around a black hole \citep{fishbone:1976}.
Two parameters describe the initial torus size:
the radius of pressure maximum $r_{max}=15.05 \rg$ and its inner edge
$r_{in}=6.4 \rg$ where $\rg$ is the gravitational radius of the
black hole of mass $M$. The torus is seeded with poloidial magnetic fields
which geometry is described by vector potential $A_\phi \sim \rho^2
r^4$, where $\rho$ is the rest-mass density of plasma and $r$ is a
radial distance from the center of the coordinate system. 
$A_\phi$ sets the initial geometry of the magnetic fields and magnetic field strength 
is specified via $\beta$ plasma parameter defined as the ratio of gas to
magnetic pressure, $\beta=P_{\rm gas}/P_{\rm mag}$. In the current model at
time $t=0$ the field is weak with maximum value of
$\beta_{max}(t=0)=100$.

The spin of the black hole\footnote{The dimensionless spin parameter
is defined as usual: $a_*\equiv Jc/GM^2$ with |$a_*<1$| where $M$ and $J$ are,
respectively, the mass and the angular momentum of a black hole.} is a
key parameter of the model that governs the shape of the emission spectrum (e.g., \citealt{moscibrodzka:2009}).
There exist constraints on spin of A0620-00 black hole
($a_*=0.12\pm0.19$) based on thin-disk continuum fitting method when the source was in
outburst in 1975 \citep{gou:2010}. Although these constraints
are model dependent, they motivate our fiducial
simulation that assumes Schwarzschild black hole.

\begin{figure*}
\begin{center}
\includegraphics[width=0.7\textwidth,angle=0]{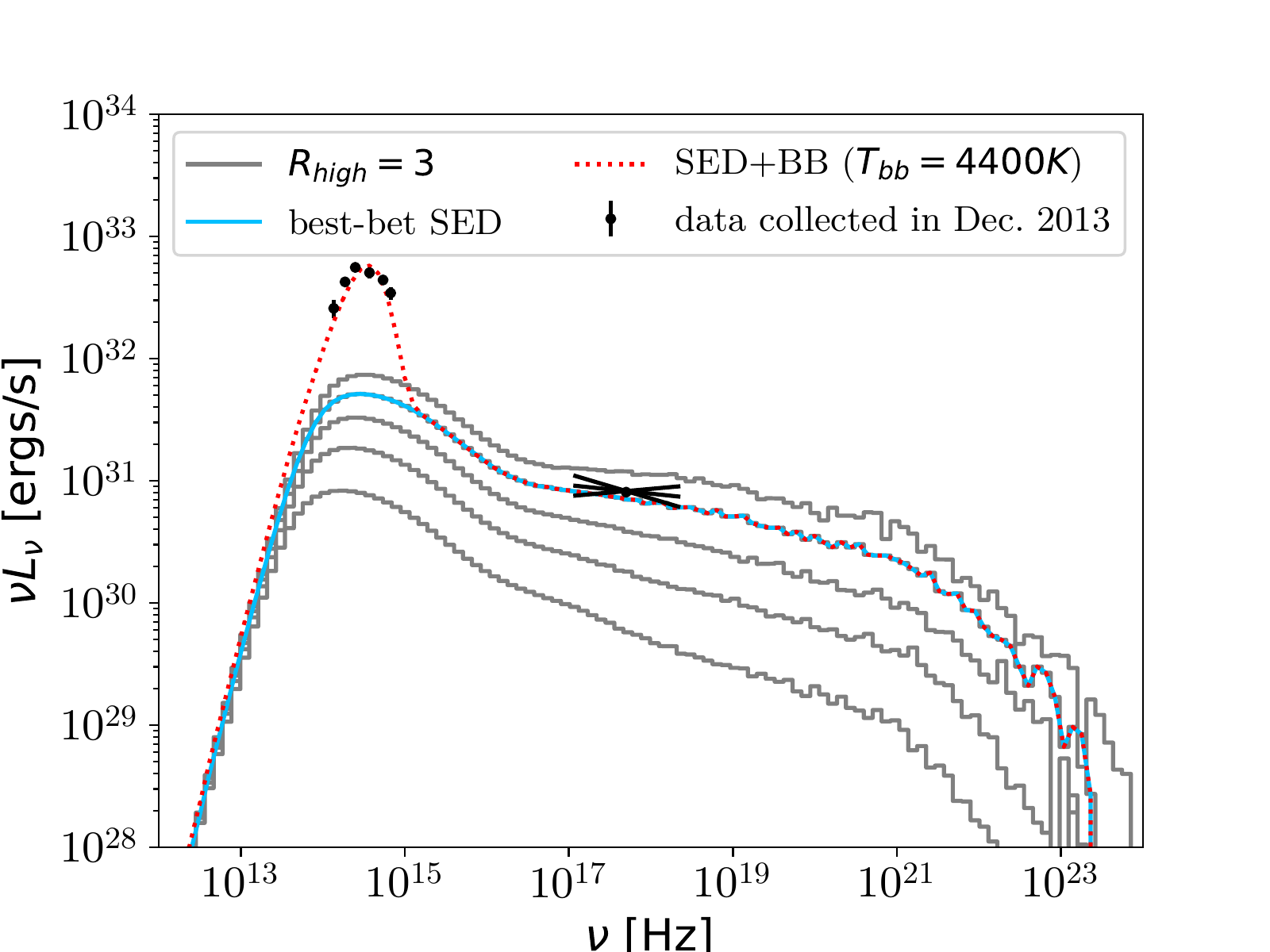}\\
\includegraphics[width=0.48\textwidth,angle=0]{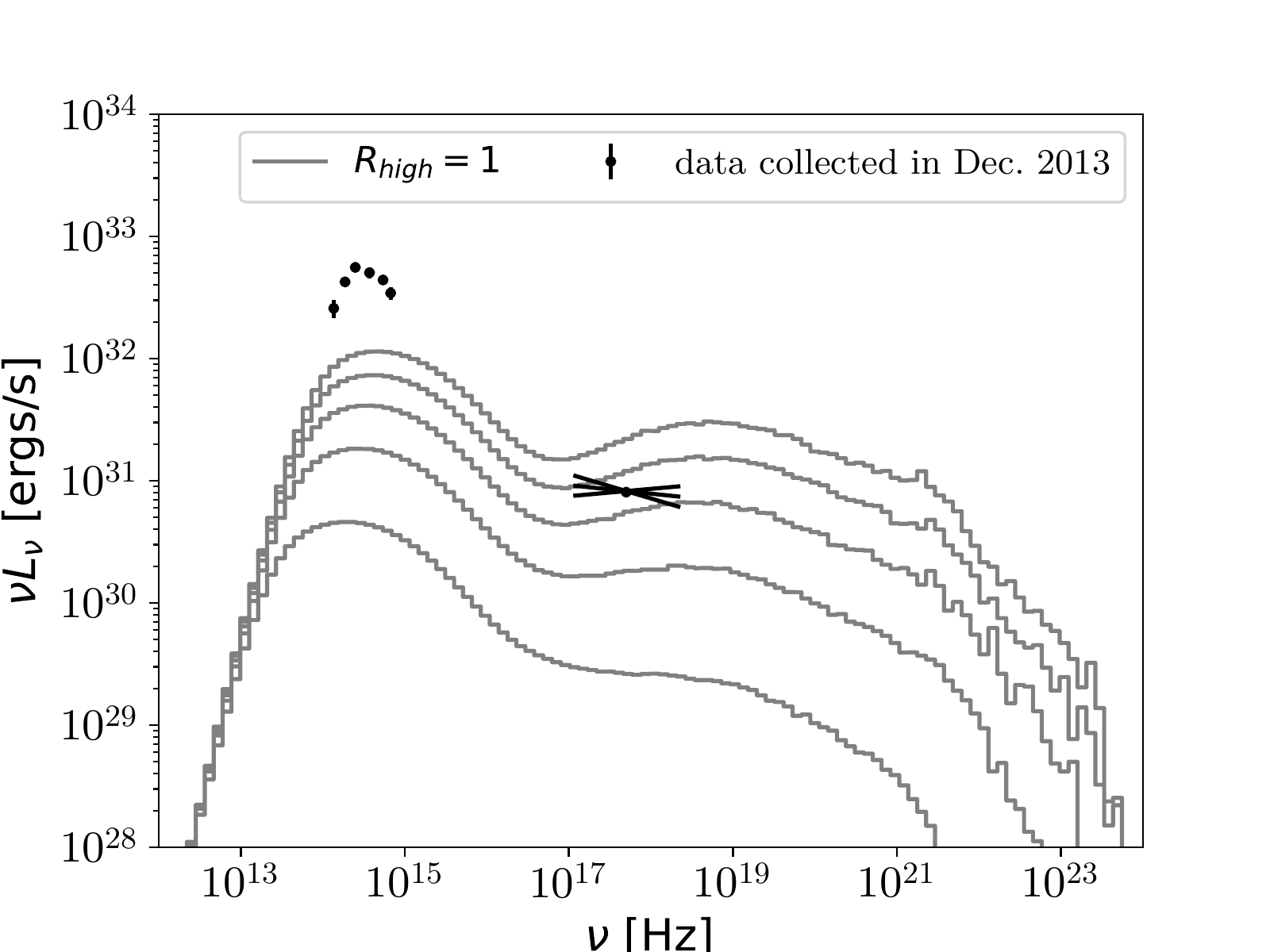}
\includegraphics[width=0.48\textwidth,angle=0]{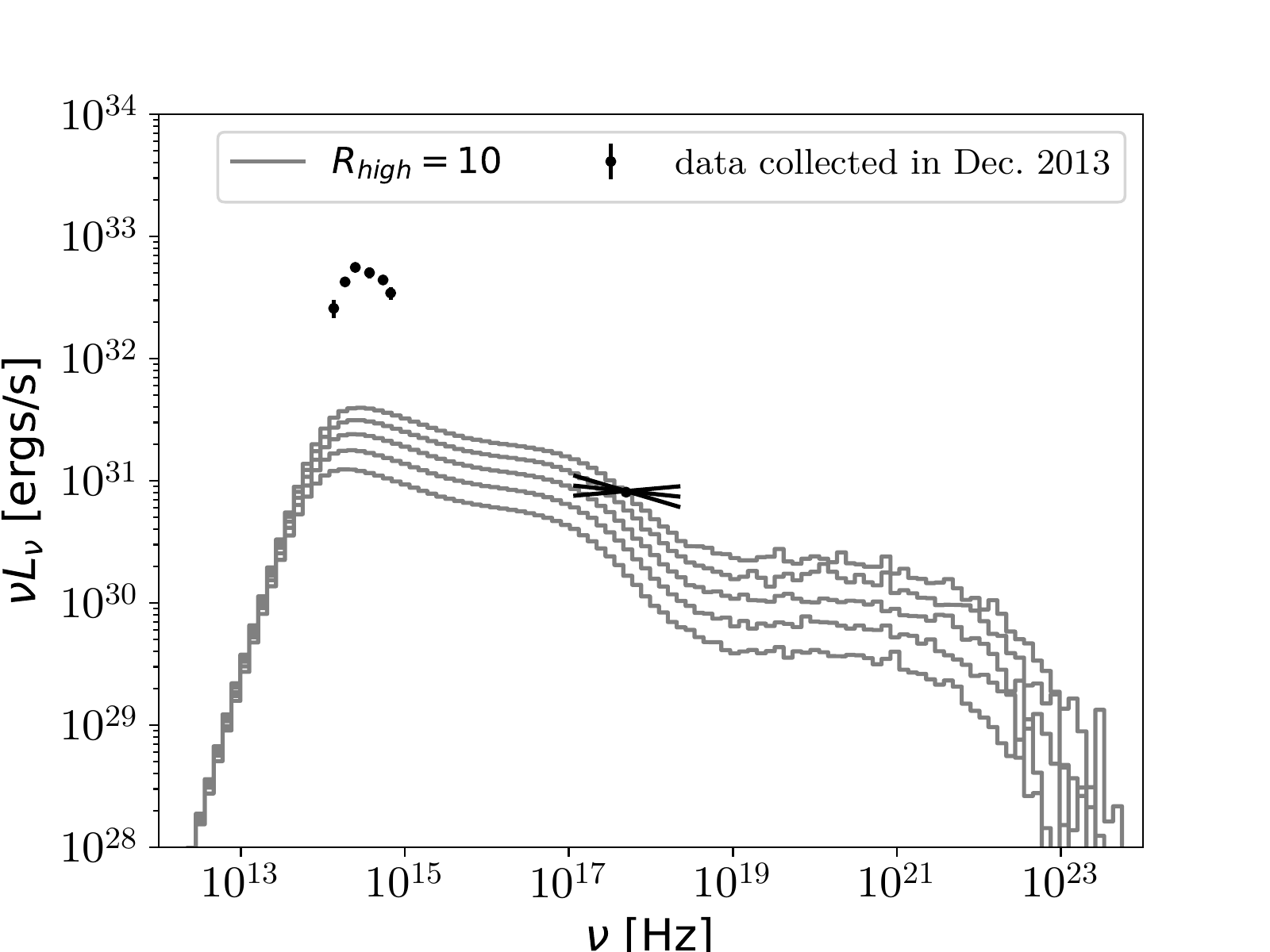}
\caption{Spectral energy distribution of synchrotron and self-synchrotron
  Compton emission produced in GRMHD simulations of hot accretion flow onto a
  stellar mass black hole in X-ray binary system.  All models assume that the
  viewing angle of the system is $\sim50$ degrees. Models with various
  electron heating assumptions ($\rhigh$ parameter) are shown in separate
  panels. For each electron temperature prescription we show spectra assuming
  a few values of $\mdot$ to find model that best matches X-ray flux and
  spectral slope. Model spectra are overplotted with observational data of
  A0620-00 system collected in 2013 \citep{dincer:2018}.  Model that recovers
  the X-ray emission characteristics is marked with a solid blue line.
  Best-bet model with thermal stellar spectrum added is shown as red dashed
  line.}\label{fig:sed}
\end{center}
\end{figure*}

Using public version of GRMHD code harm (harmpi), we integrate the equations of general
relativistic magnetohydrodynamics in Schwarzschild metric on
a three-dimensional, spherical-polar grid
that extends from within the event horizon of the black hole to 50 $\rg$. The
computational grid is spaced equally in log(r), in polar ($\theta$) and
azimuthal ($\phi$) angle direction with moderate resolution ${\rm (N_r x
  N_\theta x N_\phi)}$=(192x192x192). The inner boundary of
  the simulation is separated from the computational domain by the event
  horizon of the black hole. At the outer radius we apply outflow
  conditions. We follow the evolution of the
torus until $t_f=3200 \tg$ i.e. until the accretion flow settles down to a
steady-state phase. In the steady-state phase, the accretion flow near the
  equatorial plane is turbulent, where the turbulence develops in the initial torus 
via the magnetorotational instability (\citealt{balbus:1998}, \citealt{gammie:2003}).
In the polar regions, we observe coherent poloidal magnetic fields that
diffused there with plasma from the initial torus.
To put the current model in context of earlier works, we calculate
the magnitude of the magnetic field flux that accumulates near the event
horizon, $\phi_B = \int B^r dA / \sqrt{\dot{M} r_g^2 c}$
(the quantity is most useful when defined in dimensionless units). Models with $\phi_B\sim 1$ are
usually refered as to standard and normal evolution (SANE) models (e.g.,
\citealt{porth:2019}) and those
with $\phi_B\sim 15$ as to magnetically arrested disks (MADs) (e.g., \citealt{sasha:2011}).
In the SANE scenario, the accretion disk is turbulent which results in small variations in accretion
rate. In the MAD scenario, one expects the accretion flow to be occasionally halted by
strong magnetic forces. MADs
also produce jets that are more powerful compared to those produced by SANEs.
We find that in our model $\phi_B \lesssim 8$ meaning that our model is somewhere in-between the two distinct regimes
of accretion. The quantity $\phi_B$ can be changed by varying the size of the
initial torus or by assigning different initial field geometry/strength,
but it cannot be easily predicted. The current setup of initial magnetic
fields is chosen arbitrarily as there is no model-independent observational
constraints on the magnetic fields in A0620-00 system.

We construct mock multiwavelenght spectral energy distributions (SEDs) based on the
accretion model using postprocessing fully polarized general relativistic radiative transfer
schemes (\citealt{noble:2007,dolence:2009,moscibrodzka:2018}, and Moscibrodzka
in prep.). The radiative processes included are: synchrotron emission 
and self-synchrotron Compton (SSC). 

The primary parameter of our radiative transfer model is the distribution
function of radiating electrons. At low accretion rates the timescale for
thermal coupling between electrons and ions is much longer compared to the
dynamical timescale of the flow leading to two-temperature plasma
\citep{mahadevan:1997}. Moreover, electron acceleration may produce a
power-law distribution function. Predicting radiative characteristics of our
simulation from first principles would require a sub-grid prescription for the
evolution of electron distribution function
(\citealt{ressler:2015,ryan:2017,rowan:2017,chael:2018}).  In the present work
we adopt a simpler approach. We assume that electrons in our simulations have
relativistic, thermal (Maxwell-J{\"u}ttner) distribution function and that the
plasma has a two-temperature structure in which electron and ions temperature
coupling depends on local plasma magnetization ($\beta \equiv P_{\rm
  gas}/P_{\rm mag}$). This approach is motivated by particle-in-cell
simulations of collisionless plasma \citep{kawazura:2018}.  Following
\citealt{moscibrodzka:2016}, we calculate $T_{\rm e}$ from 
formulae that describes coupling of ion-to-electron temperatures: $T_{\rm i}/T_{\rm e} \equiv R_{\rm high}
\beta^2/(1+\beta^2) + 1/(1+\beta^2)$ where $T_{\rm i}$ is the ion
temperature followed in the GRMHD model and where $\rhigh$ is the free
parameter of the model that is constrained by fitting the model SED to the
observed spectrum. Notice that by increasing $\rhigh$ the decoupling of
electron and ions in the weakly magnetized plasma is stronger and electrons
may become subrelativistic. 

\section{Scaling GRMHD simulation to A0620-00}\label{sect:results}

The procedure of scaling GRMHD simulation to black hole in X-ray binary is
similar to the modeling emission from supermassive black holes: Sgr~A* or M87
(\citealt{moscibrodzka:2009,moscibrodzka:2016}).  Given black hole mass
$\mbh=6.6 \msun$, viewing angle $i=51$deg and the electron model, we fit model
SED to the observational data by adjusting the accretion rate onto the black
hole, $\mdot$.  However, instead of fitting emission at mm-waves (as it is
usually done in the case of these Sgr~A* and M87 because the sources are best
resolved in mm-waves) we find the model free parameters by fitting the model
spectrum to X-ray flux and X-ray spectral slope. It should be understood that
here we assume that the X-ray emission is produced by SSC on thermal
relativistic electrons in the accretion inflow/outflow, which in generally
does not have to be the case (see e.g., \citealt{connors:2017} for alternative
scenario where X-rays are produced directly by synchrotron emission from
accelerated particles).  In this work, we also do not model radio/mm emission
from the large scale jet because our simulations are designed for modeling
near horizon emission. The observed emission in GHz and lower frequencies, due
to self-absorption effects, is expected to be produced at the distances $r >
10^3 \rg$ from the black hole (\citealt{bk:1979}).

Although GRMHD simulations are intrinsically variable, here we do not model
variability of the source.  The duration of entire GRMHD simulation is $3200
\tg \approx 0.1$ second, significantly shorter compared to 7.7h orbital
period.  Hence, we pick a few snapshots at the end of the simulation time and
assume that they are representing average state of the accretion flow in the
system.

\section{Results}\label{sect:moreresults}

\begin{figure}
\begin{center}
  \includegraphics[width=0.49\textwidth,angle=0]{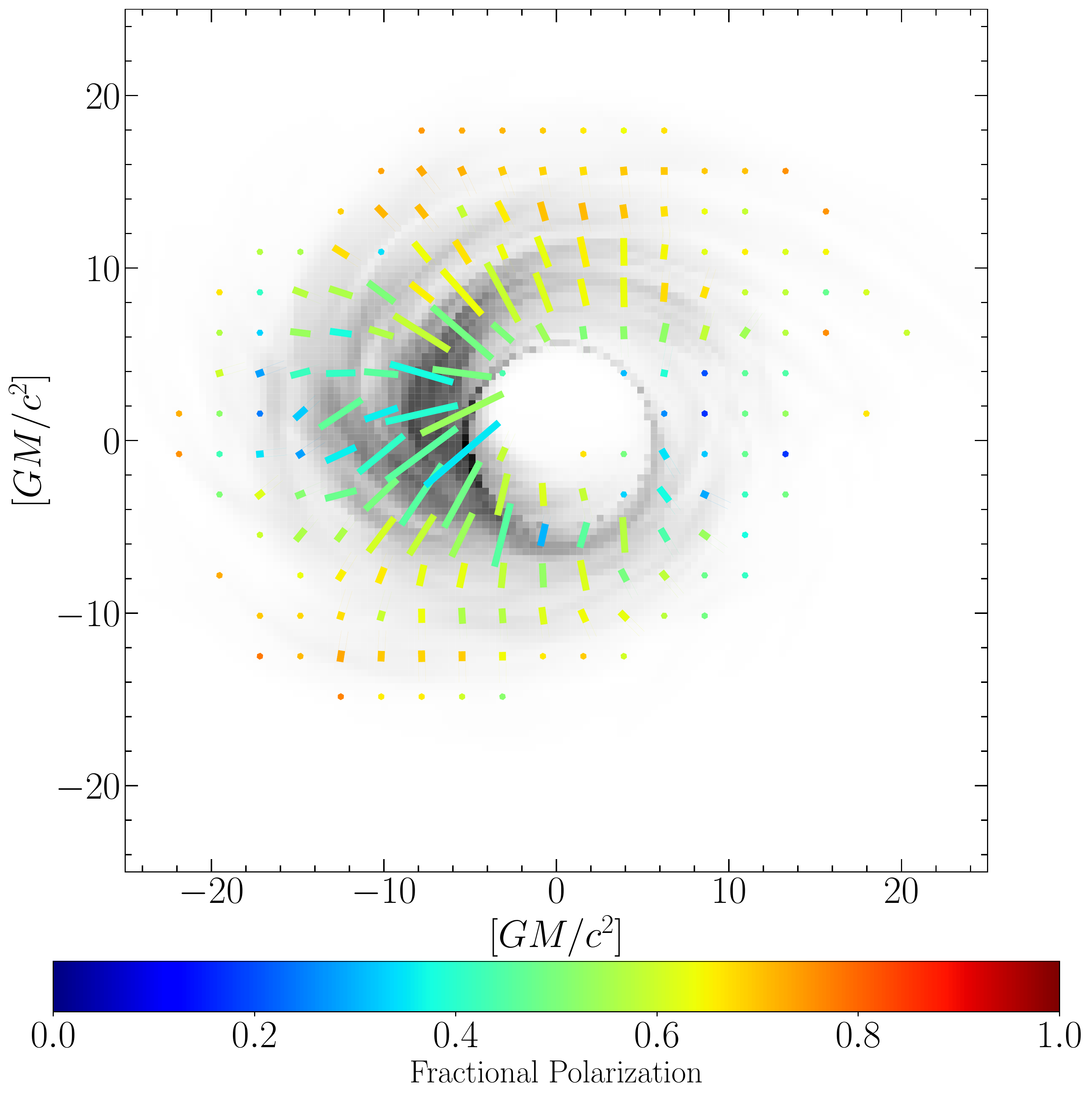}
  \caption{NIR ($K_s$ band) resolved image of the GRMHD simulation scaled to A0620-00 system. Total
    intensity is shown in gray scale (we mask regions with <1 per cent of
    maximum intensity). The color ticks show polarization of the model: the
    tick length is proportional to polarized flux, the tick color is
    fractional linear polarization and tick orientation is EVPA. We cannot
    resolve accretion flows onto stellar mass black hole on this scales so this image is
    shown only to help understand the total polarization and EVPA in
    the system. The synchrotron emission is intrinsically highly linearly polarized and shows organized EVPAs but
    the total linear polarization is beam-depolarized to less than 10 per cent.}\label{fig:map}
\end{center}
\end{figure}

In Fig.~\ref{fig:sed} we show our model SEDs overploted with the black hole
X-ray binary observational data collected in 2013,
  published in \citet{dincer:2018}. The X-ray data in Fig.~\ref{fig:sed} is {\it Chandra}
  observation no. 14656. Notice that prior to 2013 the X-ray luminosity of
  the system was weaker by a factor of 2 (in 2005)
  and 6 (in 2000) but the spectral slope has not changed significantly from
  2000 to 2013. This issue is further discussed in Sect.~\ref{sect:discussion}.

In Fig.~\ref{fig:sed}, we show SEDs for three values of $\rhigh=1,3$ and 10 and a few values of accretion
rates $\mdotedd$ for each $\rhigh$
($\mdotedd = 4 \times10^{-7} - 2 \times 10^{-6}$ for $\rhigh=1$;
$\mdotedd = 8\times10^{-7} - 2.5 \times 10^{-6}$ for $\rhigh=3$;
and $\mdotedd = 2.5 \times10^{-6} - 4 \times 10^{-6}$ for $\rhigh=10$,
where $\mdotedd=\mdot/\mdotEdd$ and $\mdotEdd=1.38 \times  10^{-7} \mdotu$).

In Fig~\ref{fig:sed}, we show that for a given electron model the changes of
accretion rate shifts the spectrum in vertical direction. The X-ray spectral slope
however is sensitive to the adopted electron heating scenario. As expected
increasing $\rhigh$ (cooler electrons in the disk plane) results in softer
X-ray spectra. Within our grid of models, the model that best describes the X-ray
luminosity and the X-ray spectral slope is the one with $R_{\rm high}=3$ and
$\mdotedd=2 \times 10^{-6}$. The best-bet
model is marked in Fig.~\ref{fig:sed} with a blue solid line.  All models with
$R_{\rm high}=1$ the spectral slope in X-ray is positive and for $R_{\rm
  high}=10$ it becomes too steep compared to observations.

The observed NIR/optical emission is dominated by light from the stellar
companion which is not included in the simulation. We model this spectral component 
with the Planck function, parameterized by a black-body temperature
$T_{\rm bb}$, integrated over the surface of the star with radius
$R_*=0.75R_\odot$. This simple approach does not include effects of deformation of the
star by tidal forces or orbital phase of the star. 
In the best-bet model the non-thermal emission from hot accretion flow produces $\sim 10$ per cent of the optical/NIR flux, 
the thermal emission from the stellar atmosphere requires $T_{\rm bb}=4400$ K.

\begin{figure*}
\begin{center}
  \includegraphics[width=0.32\textwidth,angle=0]{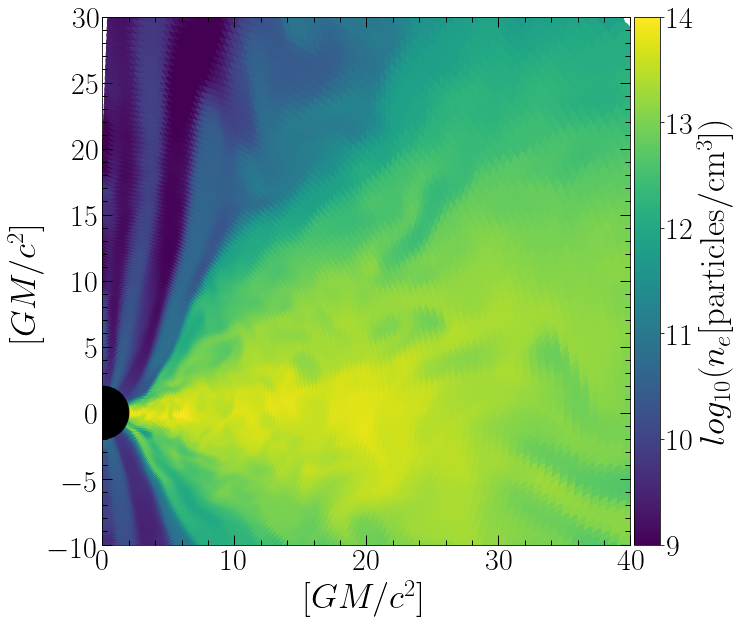}
  \includegraphics[width=0.32\textwidth,angle=0]{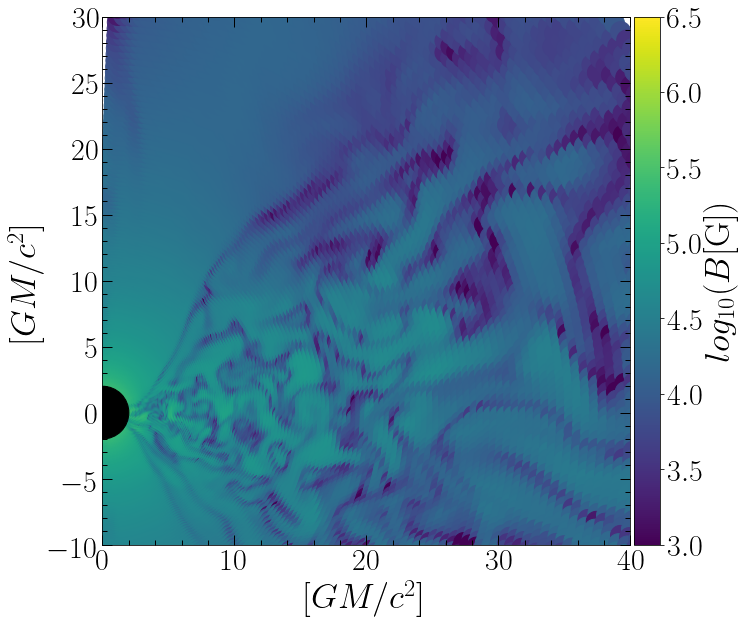}
  \includegraphics[width=0.32\textwidth,angle=0]{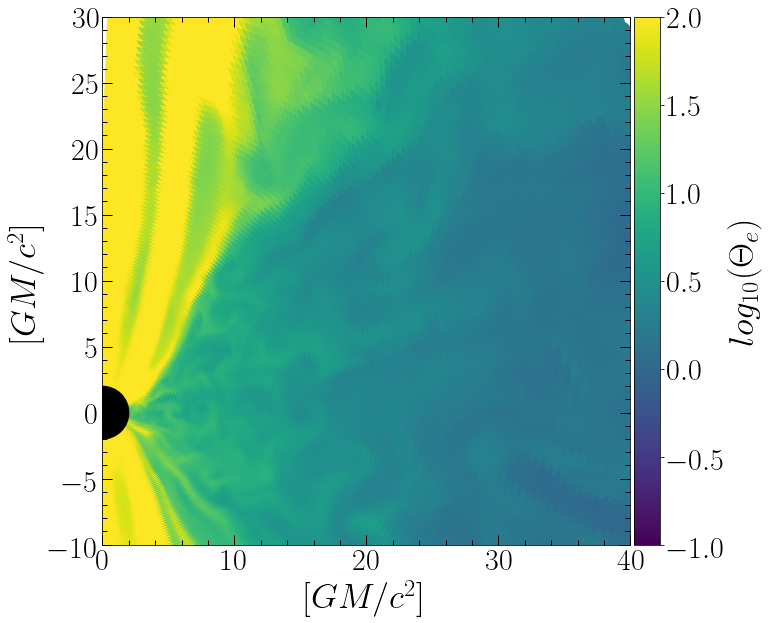}
  \caption{Maps of particle number density, magnetic field strengths and
    electron temperatures (in units of electron rest mass) in the model scaled
    to A0620-00 system. The black circle marks the event horizon of a
    Schwarzschild black hole. The plasma density and strength of magnetic
    fields near the event horizon are roughly consistent with those estimated
    by
    \citet{connors:2017} based on various semi-analytic
    jet models with simplified geometries.}\label{fig:rho_B_thetae}
\end{center}
\end{figure*}

The radiation models are fully polarized so we can report the
polarization of our best-bet model in NIR/optical band and compare it to
A0620-00 polarimetric data found in the literature \citep{russell:2016}.
Accretion rate controls optical thickness of the model, $\tau_\nu$. In our best-bet model,
$\tau_{\rm NIR/optical} \sim 1$ and it is produced near the event horizon of
the black hole. The polarization of light from an optically thin regions does
not have to be strongly polarized as the plasma can still be Faraday thick due
to some cold electrons present in the disk \citep{moscibrodzka:2017} or it can
be simply beam depolarized.
In our best model, the total fractional linear polarization ($m \equiv
\sqrt{Q_{tot}^2+U_{tot}^2}/I_{tot}$,
where $I,Q$, and $U$ are the Stokes
parameters and subscript 'tot' mean image integrated, see next section and Fig.~\ref{fig:map})
modeled in $K_s$, H, J
and Z bands are $m=9.3, 9.11, 8.9, 8.9$ per cent and in V, B, and
U filters $m=9, 9.5$, and $10$ per cent.

If we assume that the stellar emission is unpolarized and that the non-thermal
emission is a fraction $f$ of the total opt/NIR flux, then the total observed
fractional polarization is $m_{\rm obs} \equiv m f/(1+f)$.
Our model predicts $f=0.1$ which yields $m_{\rm obs} \approx 1$ per cent in both NIR and optical
bands. This predicted fractional polarization is consistent with the
observed (intrinsic, i.e., with subtracted polarization by interstellar dust)
$m \approx 1$ per cent in the NIR/optical bands \citep{russell:2016}. It is
therefore possible that the observed level of polarization is produced by a hot
accretion flow onto the black hole but this low polarization does not have to
indicate tangled/chaotic magnetic fields as suggested by \citet{russell:2016}. As evident in Fig.~\ref{fig:map},
at the assumed viewing angle of 51 deg the resolved EVPAs
(where $EVPA \equiv 0.5 arg(Q+iU$)
are approximately radial and track the toroidal
component of magnetic fields in the accretion disk.
The total (i.e., image integrated) position angles
($EVPA_{tot} \equiv 0.5 arg(Q_{tot}+iU_{tot})$ and we use convention in which EVPA=90
deg would be in horizontal position) change from 174 deg in NIR to 133 deg in
optical. Interestingly, our best-bet model also predicts that total
fractional circular polarization should vary from -1.65 percent in NIR to
-0.25 percent in optical window. In models, the circular polarization is produced by
the Faraday conversion by relativistic electrons. 

The mass accretion rate estimated from our model fitting is
$\mdotedd=2 \times 10^{-6}$ or $\dot{M}=3\times10^{-13} \mdotu$. What are
the typical densities and magnetic field strength in models scaled to this
accretion rate? In Fig.~\ref{fig:rho_B_thetae}, we show maps of particle number density,
magnetic fields strength, and electron temperatures in the GRMHD used to
produce the best-bet SED. Near horizon the $n_{\rm e} \approx 10^{14} {\rm
  [particles/cm^3]}$ and $B \approx 10^6 {\rm [G]}$. These numbers are consistent with those estimated 
  by \citet{connors:2017} based on modeling observational data from 2005.

\section{Discussion}\label{sect:discussion}

In this paper, we model the non-thermal (synchrotron and SSC) emission from
accreting black hole in X-ray binary system using GRMHD models of
accretion flows. This work focuses on the extremely sub-Eddington system
A0620-00 -- one of the few systems where the weakly radiating GRMHD models are applicable
(for similar emission models but in more luminous
sources in ``hard/low'' state see \citealt{oriordan:2016}).  In our analyses,
we model X-ray luminosity, X-ray spectral slope and, for the first time, also
NIR/optical polarization of the non-thermal emission to constrain the electron heating model
and accretion rate onto the black hole.

We find that to recover X-ray flux and the X-ray spectral slope rather strong
coupling between ions and electrons in the accretion disk ($\rhigh=3$) is
favored. The corresponding mass accretion rate is $\dot{M}=3\times10^{-13}
\mdotu$. For these parameters, our radiative model predicts
levels of observed fractional linear polarization that are consisitent with
those observed in NIR/optical bands.
Since the model X-ray spectral slope weakly depends on $\dot{M}$ and
strongly depends on $R_{\rm high}$, we expect that fitting model SEDs to X-ray
data from 2000/2005 would not change our conclusion regarding the favored
value of $R_{\rm high}$. However, we expect that model with mass accretion rate 2-3
times lower than $\dot{M}=3 \times 10^{-13} \mdotu$ would be required to match
six and two times weaker X-ray emission detected in 2000 and 2005, respectively.

In the current model, the synchrotron emission peaks in NIR/optical band with
flux $F_{\rm peak} = 260 \, {\rm [\mu Jy]}$. At $\nu <
  \nu_{\rm peak}$, the model spectrum steeply declines only due to small size
of our simulation. Although our simulation does not include a large scale jet
- the radio observations of A0620-00 still constrain some properties of an
eventual jet model. Given that the observed radio
  emission is at the level of $F_{10 GHz} \approx 15-50 \, {\rm [\mu Jy]}$
\citep{gallo:2019}, any extended jet originating from our
magnetized disk model should produce a spectrum $F_{\nu} \sim \nu^{-p}$ with
slightly inverted slope $p=0.3-0.17$ to match the observations.

An independent approach to constrain the strength of magnetic fields in the
accretion flow models in binary systems is by measuring how much angular
momentum is lost from the system via magnetized outflows and estimating the
corresponding orbital decay in the system due to magnetic braking. The orbital
decay in A0620-00 is rapid, with orbital-period derivative $\dot{P} =-0.6 \,
{\rm [ms \, yr^{-1}]}$ \citep{gonzalez:2014} and it cannot be explained by the
emission of gravitational waves alone. Magnetic braking of the system is a
possible explanation for the measured $\dot{P}$ but other explanations, such as resonant interactions
between the binary and the possible circumbinary disk, has been proposed
\citep{chen:2017}. Here we can estimate the magnitude of the magnetic braking
of the system using first-principles approach. The orbital angular momentum of a binary system is $J_{\rm orb}= \mbh
M_*/(\mbh+M_*) \sqrt{G(\mbh+M_*) d}$ where $d$ is a separation between the
star and the black hole.  In GRMHD simulations the loss of total angular
momentum through the outer boundary is defined as: $\dot{J}(r_{\rm
  out},t)=\int_\theta \int_\phi T^r_\phi dA_{\theta\phi}$, where $T^r_\phi
\equiv (\rho + \gamma u + b^2) u^r u_\phi -b^r b_\phi$ is the stress-energy
tensor describing the radial flux of angular momentum. The quantity $u$ is the
internal energy of the gas, $\gamma=4/3$ is the adiabatic index, $u^\mu$ is
the four-velocity of the gas and $b^\mu$ is a four-vector which describes
magnetic field in a frame comoving with the gas. We integrate the above
formula at $r_{\rm out}=50 \rg$ over $\theta \in (0,\pi)$ and $\phi \in
(0,2\pi)$.  In our best-bet model the ratio of angular momentum flux through
the outer boundary to the orbital angular momentum is extremely low $
\dot{J}/J_{\rm orb} \approx 10^{-22}$ [1/s],
which could account for orbital period derivative of $\dot{P} = 2
\times10^{-8} {\rm [ms \, yr^{-1}]}$ only. 
Our calculations confirm that the inner highly
magnetized, rotating accretion flow (with current $\dot{M}$) alone cannot be
responsible for the rapid orbital decay observed in the system and favor the
idea of a circumbinary disk.

How does the current picture of a quiescent accretion flow onto a stellar-mass
black hole compares to the ones around supermassive black holes in e.g., Sgr~A*?
Sgr~A* synchrotron emission at millimeter wavelengths is relatively highly polarized
\citep{bower:2018} suggesting that in the quiescent state the electrons strongly couple
to ions in RIAFs/ADAFs (see also \citealt{bower:2019} for additional
constraints on electron temperatures based on the shape of Sgr~A* Terahertz
spectrum) regardless of the black hole mass. 
If the relativistic flow around the black hole was made of two-temperature
plasma containing sub-relativistic electrons ($\rhigh \gg 1$) we would observe the
depolarization the synchrotron emission via strong Faraday effects intrinsic
to the accretion flow itself \citep{moscibrodzka:2017,alejandra:2018}. Fitting
models with $\rhigh \gg 1 $ in both sources would require considering
a model for electron acceleration into a power-law function
to recover both: the X-ray spectral
slope in A0620-00 and the significant polarization of synchrotron emission in Sgr~A*.

Finally, we have considered a single specific GRMHD simulation. A wide
survey of GRMHD simulation parameter space (including different values of
black hole spin) should be carried out for both A0620-00 and Sgr~A* to
carefully test the scale-invariance paradigm. 
Ultimately, the future EHT polarimetric images of Sgr~A* (and M87) on event
horizon scales would greatly help us to constraint the magnetic field
geometries, electron distribution functions and accretion rates in these sources. 
In A0620-00 the excess polarization could be also produced in a circumbinary disk
illuminated by both companion star and accretion flow \citep{muno:2006}. The
latter additionally constrains the electron heating models in RIAFs/ADAFs and should be
taken into account in the future works.

\section*{Acknowledgements}
This research has made use of NASA's Astrophysics Data System Bibliographic Services. I thank Sasha Tchekhovskoy for his comments.




\bibliographystyle{mnras} 
\bibliography{local} 




\bsp
\label{lastpage}
\end{document}